# Manipulating ferroelectric domains in nanostructures under electron beams

R. Ahluwalia[1], N. Ng[1], A. Schilling[2], R.G.P. McQuaid[2], D. M. Evans[2], J.M. Gregg[2], D.J. Srolovitz[3], J. F. Scott[4]

[1]*Institute of High Performance Computing, Singapore 138632*

[2]*School of Mathematics and Physics, Queen's University Belfast, Belfast BT7 1NN, United Kingdom*

[3]*Departments of Materials Science and Engineering & Mechanical Engineering and Applied Mechanics, University of Pennsylvania, Philadelphia, PA 19104, USA*

[4]*Department of Physics, Cavendish Laboratory, J. J. Thompson Avenue, Cambridge CB3 0HE, United Kingdom*

Free standing $BaTiO_3$ nanodots exhibit domain structures characterized by distinct quadrants of ferroelastic 90˚ domains in transmission electron microscopy (TEM) observations. These differ significantly from flux-closure domain patterns in the same systems imaged by piezoresponse force microscopy (PFM). Based upon a series of phase field simulations of $BaTiO_3$ nanodots, we suggest that the TEM patterns result from a radial electric field arising from electron beam charging of the nanodot. For sufficiently large charging, this converts flux-closure domain patterns to quadrant patterns with radial net polarizations. Not only does this explain the puzzling patterns that have been observed in TEM studies of ferroelectric nanodots, but also suggests how to manipulate ferroelectric domain patterns via electron beams.

The current trend of device miniaturization has led to a surge of interest in ferroelectric nanostructures [1]. This brings a number of fundamental questions on domain structure in nanoscale ferroelectric systems to the fore; these are related to the effects of free surfaces on long-range elastic and electrostatic interactions. For example, first principles calculations of nanodots suggest that flux-closing polarization vortex states form to minimize the depolarization energy arising from the bound charges at the free surfaces [2,3]. Such vortex states, involving continuous rotation of the polar vector at the dipole level, would differ profoundly from the topology of domains and domain walls classically observed in bulk crystals and thin films. While vortex states have not yet been clearly seen in experiments, the domain structures that do form in nanoscale ferroelectrics can be geometrically complex and unexpectedly sensitive to boundary conditions (e.g., exhibiting radial symmetry with Bessel function-like patterns) [4,5]. Interestingly, flux-closing domain patterns (recognized as the precursor to true vortex states [6]), can also appear naturally in some ferroelectric materials [7,8] where they are seen to take on the form of a quadrant domain arrangement. In other materials their exact configurations depend strongly upon electrical boundary conditions and electrode shape [9-11] and have also been demonstrated by domain manipulation under ambient conditions in thin films [12,13]. Such patterns also exhibit domain vertices (intersections between two or more domain walls) which can be induced and moved by application of electric fields [14,15]. Recently, material fabrication and electron microscopy techniques have reached a level where theoretical predictions about the nature of ferroelectric order at the smallest length scales is being tested experimentally [16-19]. Understanding how electron beam imaging may influence these systems is therefore extremely important. Electron beams have also been used to write domains in ferroelectric thin films [20], exploiting such an effect to engineer domain patterns could constitute a powerful technique.

Although modeling has provided some insights on domain patterns in small scale systems [2, 3, 21, 22], several fundamental puzzles remain. First, the domain patterns reported for $BaTiO_3$ (BTO) from transmission electron microscopy (TEM) studies differ unexpectedly, and strongly, from those obtained via piezoresponse force microscopy (PFM), for reasons heretofore unknown. Second, the sample diameter dependence of domain patterns provide hints [3,4,23,24] that both electric fields at the perimeter and strains may play important roles. In this letter, we examine these effects and explain the origin of the unusual patterns that have been observed in recent TEM studies of domains in ferroelectric nanodots [22]. The present study also sheds light on the mechanisms due to which the domain patterns observed in PFM and TEM can differ from each other.

BTO samples were prepared for TEM examination by cutting lamellae ($\{100\}_{pseudocubic}$ faced platelets with edges parallel to $<100>_{pc}$ measuring ~0.1 x 10 x 8 $\mu m^2$) from bulk single crystals using a Focused Ion Beam Microscope (FIB) and placing them onto (conventional) TEM grids. Once on the grid, the lamella was subsequently patterned with the FIB beam perpendicular to the broad lamella face. The resultant specimens had lateral sizes in the 0.05 – 10 μm range and shapes that included wires, disks, squares, dots, and rings [22, 25, 26]. These samples were annealed above the Curie temperature ($T_C$ = 393 K) and cooled to room temperature to form (equilibrium) domain patterns. The annealing process was performed *in-situ* (within the TEM) by rastering the sample under a focused electron beam to create a homogenous thermal profile across the entire sample [22,25,26]. Figure 1 shows the evolution of the domain configuration in a free-standing single crystal BaTiO$_3$ dot. Initially, the ferroelectric domains were in poorly organized stripes. With annealing, a quadrant configuration (four sets of 90º domains) is formed and remains stable. The boundaries between sets of 90º domains are parallel to {100}$_{pseudocubic}$ planes while the boundaries between stripe domains within each set exhibit {110}$_{pseudocubic}$ planes. This indicates an $a_1a_2$-domain pattern, where the polarization in each stripe is completely in-plane. Since these sets of 90º domains have net polarizations which are normal to the ferroelastic domain walls, it is clear that these are not flux-closure patterns. The net in-plane polarization must point into, or away from, the centre of the nanodot (described as 'center-type' domains in [27]). Interestingly, such quadrant domain configurations are ubiquitous in nano bars, squares, dots [25,26], disks and rhombohedra [Figure 2]. We consider two possible domain configurations which can correspond to the patterns shown in figures 1 and 2. In this letter, we present arguments over which of these configurations: a quadrupole pattern [Figure 3(a)] or a radial pattern [Fig. 3(b)] can be attributed to the patterns observed in TEM.

Whilst the quadrant patterns observed in the TEM nanostructures are virtually omnipresent, the inferred polarization configurations are never flux-closing in contrast to the domain patterns observed in ambient PFM (a polarization sensitive atomic force microscopy technique) studies. There are differences in sample preparation: once a lamellae intended for PFM measurements (~300 nm thick) has been cut from a bulk single crystal using a FIB, it is placed onto a Pt-coated MgO substrate with a pre-patterned electrode. The specimen is annealed at 700 $^o$C for 1 hour to improve adherence to the substrate and acid etched (2.8 mol/l HCl) to remove the Ga contamination introduced by the FIB processing. Figure 4 shows ambient domain mapping of flux-closure patterns observed via PFM, where an ac biased scan probe is rastered across the surface of the specimen. Operated in 'lateral mode', domain contrast [Figure 4(a)] originates from sensing the amplitude of induced electromechanical distortions from domains exhibiting polarization components that are oriented in the plane of the surface and perpendicular to the cantilever axis. In Figure 4(b) regions of opposite phase contrast indicate domain regions where the measured in-plane polarization component reverses by 180°. Unlike in the TEM nanodot studies, the domain stripes within each quadrant are *ac*-type domain stripes (i.e. the polarization alternates, stripe-by-stripe, from being in-plane to out-of-plane with respect to the surface). Combining these measured PFM data with knowledge of the lamella crystal orientation and that in-plane stripe polarization is oriented perpendicular to the ferroelastic domain walls (for boundary charge neutrality) shows that the in-plane polarization component of each quadrant is as indicated in Figure 4b. Note that the net in-plane polarization forms a closed loop around a core, consistent with previous PFM mapping of the quadrant configuration [28]. Such flux-closure structures have been seen to form during relaxation after application of a saturating in-plane field pulse [28]. The development of flux-closure patterns in specimens studied using PFM can be understood in terms of residual in-plane depolarizing fields (directed across the electrodes in the plane of the lamella surface). The fine-scale ferroelastic stripe domains observed within each quadrant [see Fig. 4(a)] stabilize the structure against significant disclination strains that otherwise develop in mesoscale quadrant geometries [29].

The differences between the PFM and TEM images for $BaTiO_3$ are remarkable. While both domain patterns consist of four sets of $90^o$ domains (i.e. quadrants), the main difference between PFM and TEM images is in the direction of the net polar orientation in each quadrant. In this paper we would like to address possible factors which can be responsible for these pronounced differences in the observed TEM and PFM patterns in the same material. Although the specimens are of somewhat different sizes (100 nm and 300 nm in thickness, and half and several microns in lateral size, in the TEM and PFM specimens, respectively), and that the measurements are made in somewhat different environments (vacuum and air for the TEM and PFM observations, respectively), it is surprising how different the polarization patterns are.

While the flux-closure patterns in PFM images can be rationalized in terms of simple depolarizing fields developed across two opposing lateral edges of the lamella, the formation of non-flux-closing quadrant domain patterns in TEM specimens remains a mystery. This is especially so since the thermodynamic drive to minimize depolarizing fields is expected to be larger in the TEM nanodot geometry. It has been suggested that such patterns could arise from quadrupolar electric fields [26]. However, the physical origin of such fields is unknown. One important distinction however is that an electron beam is applied to the sample (only) in the TEM case. It is well known that electron beams can lead to specimen charging [30] and, for ferroelectrics specifically, can generate internal electric fields large enough to cause polarization switching [30]. In this paper we emphasize that electron beam charging may be responsible for quadrant pattern formation, and furthermore, that the resulting polarization profile associated with this pattern is radial, rather than quadrupolar. We investigate this proposal via real space phase field simulations of free standing nanostructures (with uniform free charge densities).

The phase field model provides a powerful technique to simulate domain structures in ferroelectrics [31, 32]. The method is based on the Landau-Ginzburg free energy of a ferroelectric material which incorporates elastic and electrostatic effects. The model used is the same as that in [21], except the effect of free charge that is included in the present calculations. The details of the model are provided in the supplementary materials.

In the 2D simulations reported here, we consider a mechanically unconstrained nanodot with traction free boundary conditions on all lateral surfaces (i.e, $\sigma_{ij} \cdot n_j = 0$, where $\vec{n}$ is the unit surface normal). $P_x = P_y = 0$ in the vacuum outside the nanodot. The electrostatic potential satisfies the boundary condition $\vec{\nabla}\phi \cdot \vec{n} = 0$ on the nanodot surface. This model has been used to simulate domain patterns in free standing ferroelectrics samples of different sizes and geometries [21] in both charge compensated and uncompensated situations. The effect of free charge can be incorporated the model by introducing a free charge density $\rho(\vec{r})$ in the Gauss's law as $\vec{\nabla} \bullet \vec{D} = \rho(\vec{r})$. To understand the role played by a free charge density on domain patterns, we simulate domain patterns for $\rho(\vec{r}) =$ *constant* and compare it with the $\rho(\vec{r}) = 0$ case. The phase field equations are numerically integrated using a finite difference procedure and the parameters are chosen to represent $BaTiO_3$ (these constants and their normalization are given explicitly in the Supplementary Materials).

We consider free-standing nanodots where the bound charge at the free surface is not compensated. To describe trapped charge, we impose a homogeneous charge density $\rho(\vec{r}) = -eN_e$, where $e$ is the electronic charge and $N_e$ is trapped electron density. The presence of free charge layers in ferroelectric thin films can strongly influence the observed domain patterns [32]. In order to answer the question how a uniform charge density influence the domain patterns in a nanodot we performed simulations of domain patterns for different values of $N_e$ for times up to $t^* = 10^4$. We consider situations where the nanodots are quenched from the high temperature paraelectric state into the ferroelectric state ($T = 375K$) and that there is a pre-existing trapped charge density in the paraelectric state as may be anticipated in the scenario of electron beam induced field-cooling in the TEM. We also assume that the charges are immobile during the domain formation process. We chose a temperature above room temperature to avoid metastable orthorhombic domain formation. The occurrence of metastable orthorhombic domains at room temperature delays the formation of the quadrant patterns that are expected to be similar to those at $T = 375K$.

Figure 5 shows the domain patterns for the polarization components and the strains $\varepsilon_{xx} - \varepsilon_{yy}$ for the $N_e = 0$, and $N_e = 1.56 \times 10^{28}/m^3$ cases at $t^* = 10^4$ We first examine the $N_e = 0$ case. Although no large scale vortex states are observed, we do observe flux-closure domains and the bound charge $\vec{P} \cdot \vec{n} = 0$ on the surface. Inside the nanodot, we see randomly arranged ferroelastic domain bundles. However, no quadrant patterns are observed. These patterns are analogous to those observed in [21] for the case of uncompensated charges. We also simulated several cases for finite $N_e$. When $N_e = 1.56 \times 10^{28}/m^3$, a clear quadrant pattern is observed (note the net polarizations at the corners). The nanodot center exhibits a disordered pattern which avoids a divergence of the polarization vectors at the core. The polarization vectors are parallel to surface (flux-closure) in most regions for $N_e = 0$ but not when $N_e = 1.56 \times 10^{28}/m^3$ (i.e., $\vec{P} \cdot \vec{n} \neq 0$) - the bound charge at the free surface balances the free charge density. The domain pattern for the finite $N_e$ case is reminiscent of those observed in [22] and in Figs. 1 and 2. Although the simulated patterns are more disordered than in the experimental images, the formation of a quadrant configuration is clearly observed.

In order to relate simulation parameters to experimental, we note that analogous TEM experiments on graphene [33,34] found that an electron beam of ~$10^5$ electrons/(nm²s) for 30 seconds (total charge of ~3 × $10^{24}$ electrons/m²) was sufficient to induce defect-pair creation analogous to Kosterlitz-Thouless melting. Assuming a nanodot of 100 nm thickness and $N_e = 1.56 \times 10^{28}/m^3$, the number of electrons per unit area in our simulations is ~$1.56 \times 10^{21}$ electrons/m². This is smaller than the charge which was enough to cause defect pair nucleation in graphene [33] by a factor of ~2000. While this is not a sharp threshold, it gives a rough estimate of the charging required for TEM patterns to produce symmetries that differ strongly from the non-charging, PFM domain structures.

To understand the dynamics of how an uncharged sample responds to imposed charge, we performed additional simulations in which an initially uncharged pattern, with flux-closure domains, is subjected to a finite free-charge density. Figure 6 shows the evolution of the domain pattern in response to the imposed charge. The randomly aligned ferroelastic domain bundles evolve into a quadrant structure. This shows that quadrant patterns may be prepared by imposition of sufficiently strong charging, such as that occuring in an intense electron beam.

Furthermore we address the question of what drives quadrant pattern formation. As is well known, the electric field inside a uniformly charged symmetric body is radial with zero magnitude at the center (see Fig. S1 in the Supplementary Information) . This radial electric field can influence domain patterns by promoting polarisation components that point to/away from the centre of the nanodot. For $BaTiO_3$, the radial field tends to stabilize a quadrant pattern of ferroelastic domains. This can be rationalised by considering, for example, the electric field component oriented along [110], a pattern of [100]- and [010]-oriented polarizations will be stabilized resulting in a quadrant with net polarisation along [110]. Using this argument, we can schematically construct a quadrant domain pattern (see Fig. 3(b)) that is consistent with those observed experimentally (Fig 1). Note that this pattern necessarily involves charged domain walls in the core regions. However, note that the simulations show no charged walls in the core region, (see Fig. 5). Instead, a complex pattern with local flux-closure is observed. This pattern forms in response to the strong depolarizing and elastic fields that arise from the radial polarizations at the corners. This is consistent with the TEM observations in which the core regions exhibit a pattern that lacks quadrant symmetry (Figs. 1 and 2).

In the following we explore a possible origin of the charge in the nanodot. While electron microscopy is routinely employed to image ferroelectric domains, electron beam charging is common in low electrical conductivity materials. Charging during imaging may be insufficient to alter domain patterns. However, in the present $BaTiO_3$ nanodots experiments, the samples were also heated above the Curie temperature using the electron beam. The exposure to an intense electron beam could lead to significant charging and therefore to the formation of near radial electric fields. Intruguingly, whilst the $BaTiO_3$ quadrant domain patterns exhibit flux-closure in ambient PFM studies but not in TEM studies, previously studies of PZT nanodots in a TEM environment [35] revealed the expected flux-closure type quadrant patterns. The key difference is that these PZT dots were heated through the Curie temperature in air outside the TEM (BTO nanodots were heated up in situ in TEM), and therefore they were not subjected to the radial fields during the heating and cooling cycle.

We would like to highlight that although in this work only single-crystal structures are considered, similar domain patterns have been observed in thin films where domains were written using a biased scan probe tip in direct contact with the film surface. This is common practice for measuring hysteresis and writing domain patterns at the nanoscale where the tip-writing field is inhomogeneous and often seen to have radial symmetry [36]; this is well evidenced by the circular profile of domains written in $PbZrTiO_3$ [37] and the radial polarization profile of domain structures engineered in $BiFeO_3$ thin films [27]. Similar quadrant patterns have been observed in localized regions in $PbTiO_3$ thin films [38], if there are localized charged defects then the quadrant formation may be due to the mechanism proposed here. Finally, we note that whilst the probe tip-field and TEM charging scenario may both generate fields with radial field symmetry, the amplitude profile is not the same in both cases. In experiments utilizing a biased scan probe the field decays away from the probe point of contact, as opposed to increasing from the center of the structure as modeled for the TEM charging case; the effect of this specific field configuration on ferroelastic domain structure has not been well explored.

The present results explain the puzzling quadrant patterns that have been observed in TEM imaging of ferroelectric domains in nanodots and they might serve as a warning that the domain configuration under electron beam observation is a non-equilibrium property. Using phase field simulations, it is demonstated that electron beam induced radial electric fields stabilize such patterns. The electron densities required to form the quadrant patterns are significantly higher than typical defect densities in ferroelectrics, implying that external charging may be involved. We suggest that such charging occurs during electron beam heating of the sample. This implies that an electron beam may be intentionally used to "write" such patterns and suggests a possible avenue toward new classes of ferroelectric devices that utilize such electron beam engineered domain states.

FIGURES

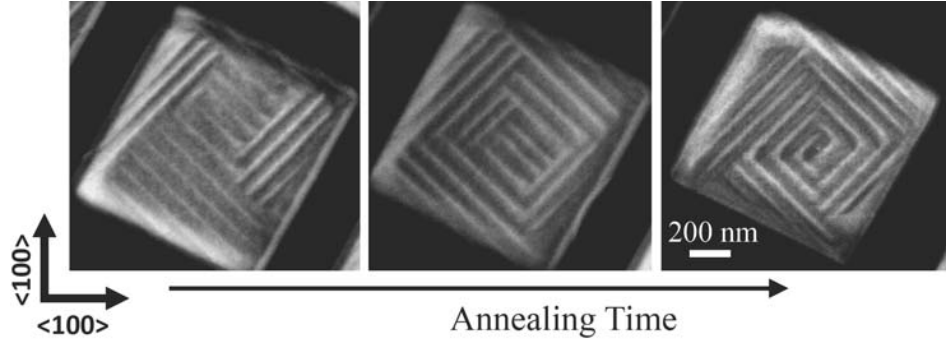

**Figure 1**. Scanning TEM imaging of the evolution of the domain pattern (over a period of approximately 2 minutes) towards a quadrant structure consisting of four sets of 90º domains.

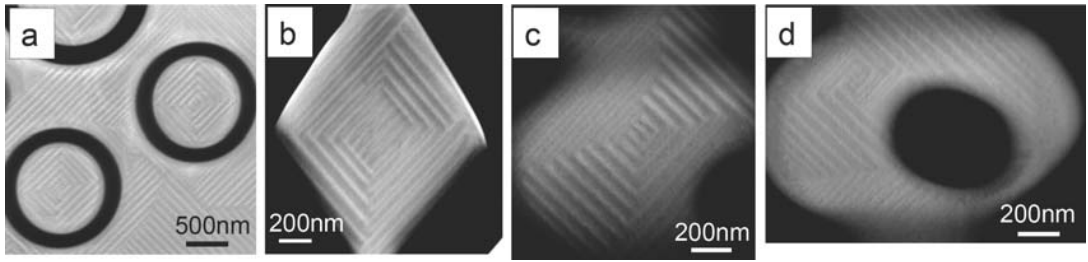

**Figure 2**. Scanning TEM images of ferroelectric quadrant domain structures in (a) disks, (b) a rhombohedra, (c) an asymmetric ring, and (d) a complex asymmetric pattern.

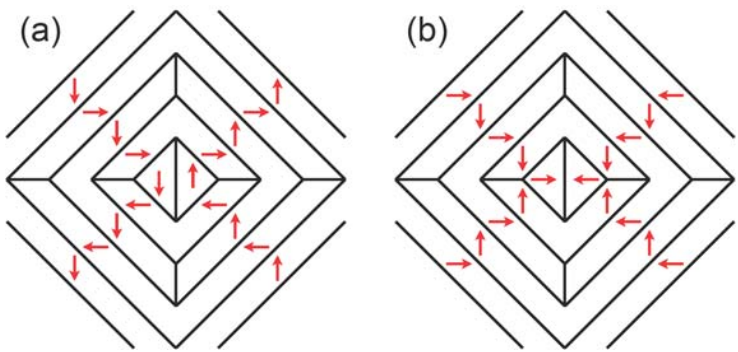

**Figure 3**. Two possible domain arrangements suggested by the TEM images showing (a) quadrupolar pattern or (b) radial pattern.

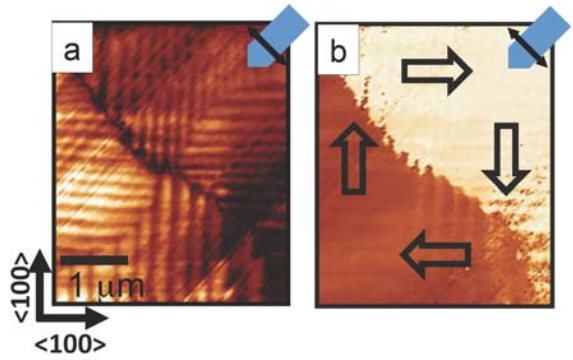

**Figure 4.** A typical flux-closing domain configuration imaged in ambient conditions in a BaTiO3 lamellae using lateral mode PFM. (a) Piezoresponse amplitude reveals quadrant domain configuration with finer stripe domains within each quadrant. Scanprobe cantilever orientation is illustrated in blue and polarisation components sensed in lateral mode are indicated by the black double-headed arrow. (b) PFM phase image can be used to identify in-plane polarisation orientations in each quadrant (black hollow arrows).

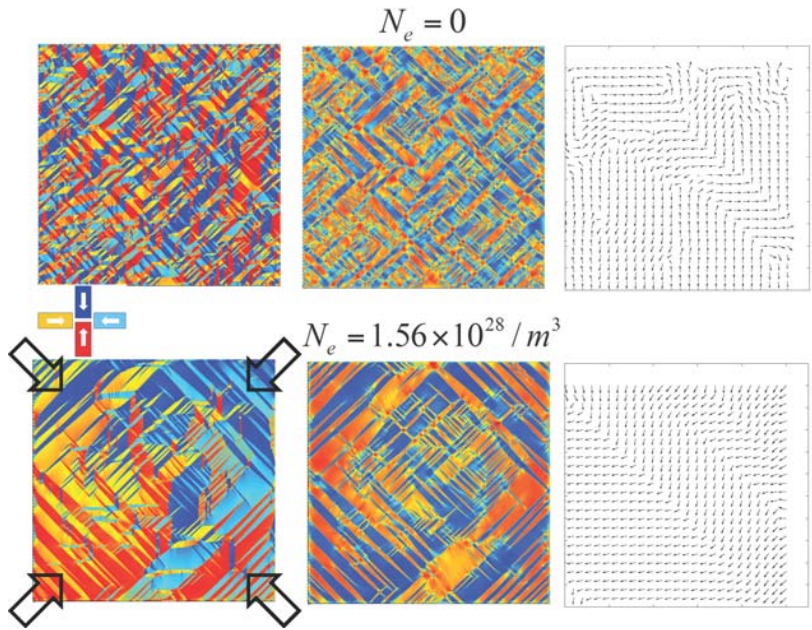

**Figure 5.** Simulated patterns without free charge (top) and with uniform free charge (bottom) for a 1000 nm × 1000 nm nanodot. The left images depict polarization domains, the center images show the associated ferroelastic domain pattern, and the right images show the polarization vectors associated with the top right corner of the images on the left.

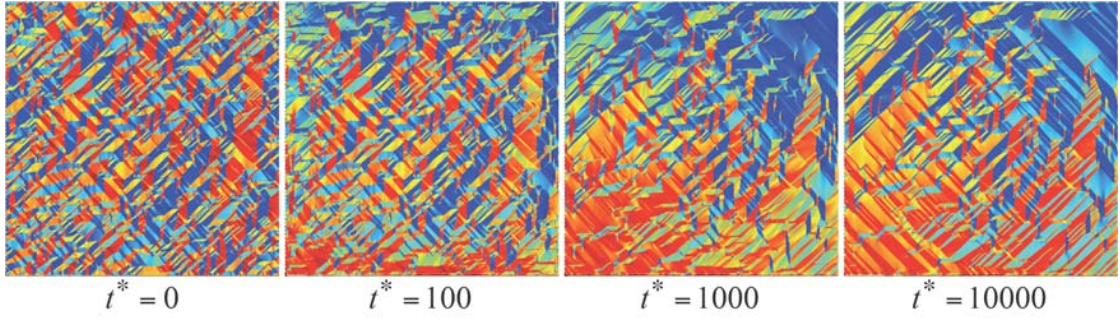

**Figure 6.** Domain evolution induced by imposition of a charge density of $N_e = 1.56 \times 10^{28}/m^3$ on an initially uncharged sample (with flux-closure domains)